%% file: paper.tex
\documentclass[conference]{IEEEtran}
\IEEEoverridecommandlockouts
\usepackage{cite}
\usepackage{amsmath,amssymb,amsfonts}
\usepackage{algorithmic}
\usepackage{graphicx}
\usepackage{textcomp}
\usepackage{xcolor}
\usepackage[final]{pdfpages}
\usepackage{tikz}
\usepackage{amsmath}
\usepackage{enumitem}
\setlist{nosep}
\usepackage{filecontents}
\usepackage{sistyle}
\SIthousandsep{,}
\usepackage{multirow}
\usepackage{tabularx}
\usepackage{subfig}
\usepackage{placeins}
\usepackage{float}
\usepackage{listings}
\usepackage{xcolor}
\usepackage{color}
\usepackage[bottom]{footmisc}
\usepackage{url}
\usepackage{seqsplit}
\PassOptionsToPackage{hyphens}{url}\usepackage{hyperref}
\newcommand{\hash}[1]{{\ttfamily\seqsplit{#1}}}
\definecolor{codegreen}{rgb}{0,0.6,0}
\definecolor{codegray}{rgb}{0.5,0.5,0.5}
\definecolor{codepurple}{rgb}{0.58,0,0.82}
\definecolor{backcolour}{rgb}{0.95,0.95,0.92}
\lstdefinestyle{mystyle}{
    backgroundcolor=\color{backcolour},   
    commentstyle=\color{codegreen},
    keywordstyle=\color{magenta},
    numberstyle=\tiny\color{codegray},
    stringstyle=\color{codepurple},
    basicstyle=\ttfamily\footnotesize,
    breakatwhitespace=false,         
    breaklines=true,                 
    captionpos=b,                    
    keepspaces=true,                 
    numbers=left,                    
    numbersep=5pt,                  
    showspaces=false,                
    showstringspaces=false,
    showtabs=false,                  
    tabsize=2
}
\def\num #1{\numA#1\empty\empty\empty#1\end}
\def\numA #1#2#3{%
   \ifx #1\empty \afterelax{\numB}\fi
   \ifx #2\empty \afterelax{\numB{}}\fi
   \ifx #3\empty \afterelax{\ea\numB\ignoreit}\fi
   \ea \numA \ignoreit \relax
}
\def\numB #1#2#3#4{#1#2#3\ifx#4\end\else \numseparator \ea\numB\ea#4\fi}

\def\afterelax#1#2\relax{\fi#1}
\def\ignoreit#1{}
\let\ea=\expandafter

\def\numseparator{,}

\usepackage[english]{babel}
\usepackage{blindtext}

\def\BibTeX{{\rm B\kern-.05em{\sc i\kern-.025em b}\kern-.08em
    T\kern-.1667em\lower.7ex\hbox{E}\kern-.125emX}}
\begin{document}

\title{Understanding the Cryptocurrency Free Giveaway Scam Disseminated on Twitter Lists}
\author{
\IEEEauthorblockN{Kai Li}
\IEEEauthorblockA{\textit{Department of Computer Science} \\
\textit{San Diego State University}\\
San Diego, USA \\
kli5@sdsu.edu}
\and
\IEEEauthorblockN{Darren Lee}
\IEEEauthorblockA{\textit{Department of Computer Science} \\
\textit{San Diego State University}\\
San Diego, USA \\
dlee0083@sdsu.edu}
\and
\IEEEauthorblockN{Shixuan Guan}
\IEEEauthorblockA{\textit{Department of Computer Science} \\
\textit{San Diego State University}\\
San Diego, USA \\
sguan4105@sdsu.edu}
}

\maketitle
\input{text/abs.tex}
\thispagestyle{plain}
\pagestyle{plain}

\input{text/intro.tex}
\input{text/background.tex}
\input{text/system.tex}
\input{text/results_lists.tex}
\input{text/results_profit.tex}
\input{text/conclusion.tex}

\bibliographystyle{IEEEtran}
\bibliography{bkc, bkc_scam, twitter_scam, youtube_scam}

\end{document}

%% file: text/abs.tex
\begin{abstract}
This paper presents a comprehensive analysis of the cryptocurrency free giveaway scam disseminated in a new distribution channel, Twitter lists. To collect and detect the scam in this channel, unlike existing scam detection systems that rely on manual effort, this paper develops a fully automated scam detection system, \textit{GiveawayScamHunter}, to continuously collect lists from Twitter and utilize a Nature-Language-Processing (NLP) model to automatically detect the free giveaway scam and extract the scam cryptocurrency address.

By running \textit{GiveawayScamHunter} from June 2022 to June 2023, we detected 95,111 free giveaway scam lists on Twitter that were created by thousands of Twitter accounts. Through analyzing the list creator accounts, our work reveals that scammers have combined different strategies to spread the scam, including compromising popular accounts and creating spam accounts on Twitter. Our analysis result shows that 43.9\% of spam accounts still remain active as of this writing. Furthermore, we collected 327 free giveaway domains and 121 new scam cryptocurrency addresses. By tracking the transactions of the scam cryptocurrency addresses, this work uncovers that over 365 victims have been attacked by the scam, resulting in an estimated financial loss of 872K USD.

Overall, this work sheds light on the tactics, scale, and impact of free giveaway scams disseminated on Twitter lists, emphasizing the urgent need for effective detection and prevention mechanisms to protect social media users from such fraudulent activity.
\end{abstract}

%% file: text/intro.tex
\section{Introduction}
\begin{figure*}[!ht]
  \centering
    \subfloat[Ethereum giveaway website]{%
  \includegraphics[width=0.375\textwidth]{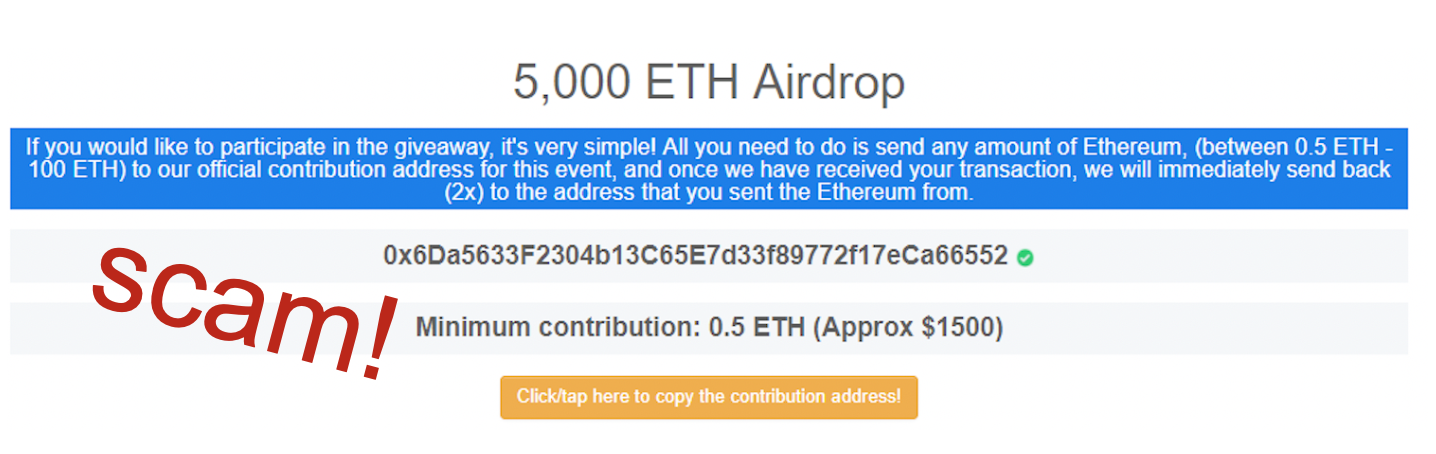}
  \label{fig:ether_scam_site}
    }%
    \subfloat[Crafted transactions]{%
  \includegraphics[width=0.375\textwidth]{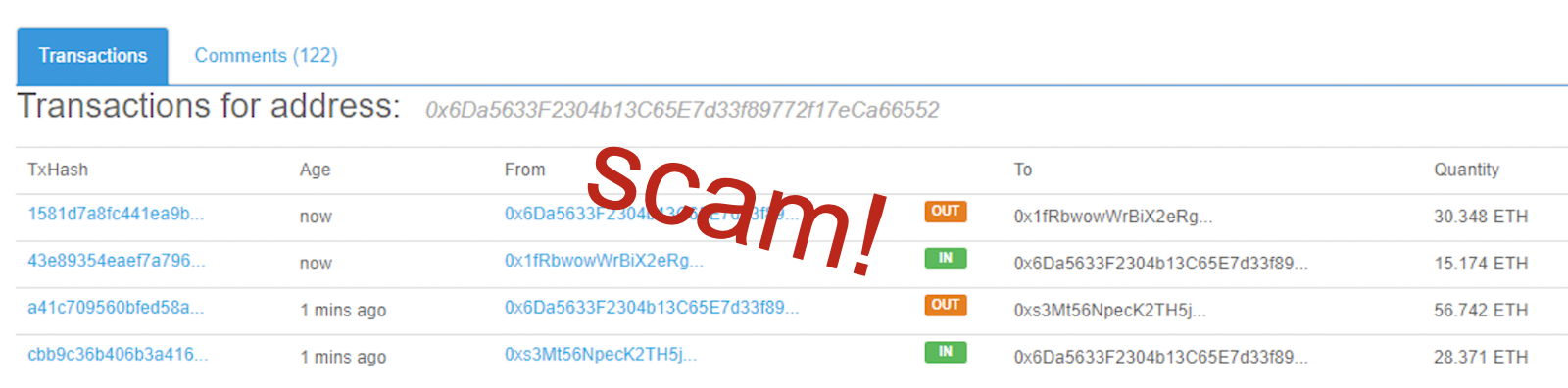}
  \label{fig:ether_scam_txs}}
    \subfloat[Created Twitter list]{%
  \includegraphics[width=0.25\textwidth]{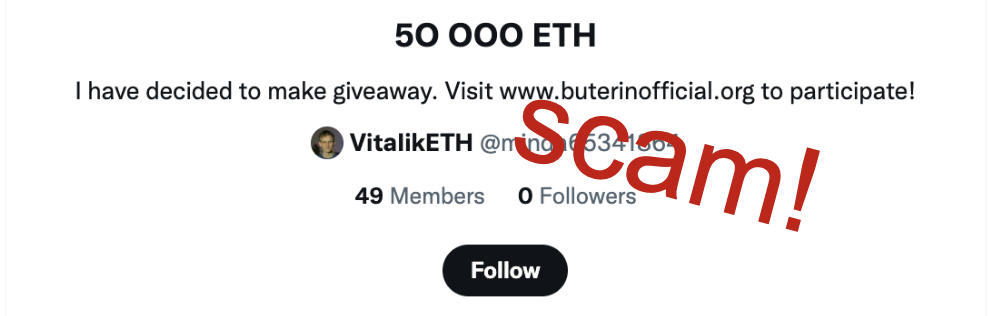}
  \label{fig:ether_scam_list}}
  \caption{A sample Ethereum giveaway scam hosted on \url{www.buterinofficial.com} and disseminated through a Twitter list.}%
  \label{fig:ether_scam}
\end{figure*}

With the flourish of cryptocurrency markets, recent years have witnessed a surge of cryptocurrency scams that play various tricks to steal funds from victims, including Ponzi schemes\cite{kell2021forsage, bian2021image, xia20covidscams, bartoletti2020dissecting, bartoletti2018data, chen2018detecting}, phishing scams\cite{chen2020phishing, badawi2020automatic}, scam tokens\cite{gao2020tracking, xia21scams}, and cryptocurrency exchange scams\cite{xia2020characterizing}. Among them, the free giveaway scam is an emerging scam that works by creating legitimate-looking websites or hacking social media accounts of celebrities to announce free giveaway schemes, claiming that anyone who transfers a certain amount of funds to them will receive double or more back, as exemplified in Figure~\ref{fig:ether_scam_site}. 

In a typical free giveaway scam announced in a legitimate-looking website, in order to increase the visibility, the scammer tend to utilize all kinds of distribution channels to spread the URL of the website, including online social networks (OSNs) such as YouTube, Twitter posts. Although existing work~\cite{xia20covidscams, vakilinia2022cryptocurrency, xigao2023doublenothing} have tackled the free giveaway scam, all of them focus on some known distribution channels or rely on heuristics to detect the free giveaway scheme, thus cannot cover scam instances spread in new channels. In addition, they all rely on manual effort to assess the collected data to identify the scam, which is tedious and cannot scale to analyze a large amount of social network data. 

\textbf{New distribution channel and detection system.} In this work, we identified a new distribution channel used by scammers in disseminating the free giveaway scam, which is the Twitter list. We found that its permission-less and pushing features have being abused by scammers to actively push the scam to Twitter users. Due to the limitations of existing work in detecting such a free giveaway scam in the new distribution channel, this work strive to complement them by a fully automated scam detection system that can discover free giveaway scam instances on Twitter lists. Our detection system named \textit{GiveawayScamHunter} can continuously collect Twitter lists and utilize a Natural Language Processing (NLP) model to detect lists containing the free giveaway scam. Further more, \textit{GiveawayScamHunter} can extract the URLs of free giveaway websites and visit them to extract the associated scam cryptocurrency addresses. 

\textbf{Detection results.} We implemented and deployed \textit{GiveawayScamHunter} from June 2022 to June 2023. In summary, \textit{GiveawayScamHunter} has detected \num{95111} free giveaway scam lists that were created from \num{87617} Twitter accounts and have reached over \num{3.7} million Twitter users. Notably, through analyzing the profiles of list creator accounts, our work reveals that the scammers have combined different strategies to spread the scam, including compromising popular accounts and registering spam accounts. Our analysis also shows that 43.9\% of spam accounts still remain active as of this writing, implying the urgent needs of detecting the spam accounts and preventing the dissemination on Twitter. Moreover, \textit{GiveawayScamHunter} has collected \num{327} free giveaway URLs and \num{121} scam cryptocurrency addresses, out of which 310 URLs and 121 addresses are not reported in existing work. The collected scam cryptocurrency addresses span over 5 blockchains including Bitcoin, Ethereum, Binance, Cardano, Ripple. By tracking the transaction history of the scam addresses, we estimate that over 365 victims have been attacked by the scam, leading to a financial loss of up to 872K USD. 

\textbf{Contributions:} we made the following contributions.
\begin{itemize}[leftmargin=*]
\itemsep0em 
 \item \textit{New distribution channel.} We identified a new distribution channel, Twitter list, which has been abused by free giveaway scammers to spread the scam due to its permission-less and pushing features, allowing scammers to actively push scam schemes to Twitter users.  
 \item \textit{New scam detection system.} We developed a fully automated scam detection system, \textit{GiveawayScamHunter}. Unlike existing work, our \textit{GiveawayScamHunter} does not rely on manual effort but a NLP model to automatically detect the free giveaway scam.
 \item \textit{New scams and spam accounts.} Our detection system \textit{GiveawayScamHunter} has collected tens of thousands of free giveaway scam lists and spam accounts on Twitter, as well as hundreds of free giveaway websites and scam cryptocurrency addresses that are not reported in existing work. We will open source our detection system and data-set to benefit future studies.
 \item \textit{New understandings.} Our work reveals the strategies of scammers in spreading the scam through Twitter lists as well as the scale of deceived victims and the financial loss, emphasizing the urgent needs of detecting and preventing the dissemination of the scam.
\end{itemize}

\noindent\textbf{Road-map:} Sec.~\ref{sec:pre} provides the necessary background of the free giveaway scam spread on Twitter lists. Sec.~\ref{sec:system} details the design and implementation of \textit{GiveawayScamHunter}. Sec.~\ref{sec:list} presents the analysis of scam lists and Sec.~\ref{sec:victim_profit} discusses the uncovered victims and financial loss. Sec.~\ref{sec:discuss} discusses the implications and limitations of this work. Sec.~\ref{sec:related} introduces related work and Sec.~\ref{sec:conclusion} concludes this paper.

%% file: text/background.tex
\begin{table*}[!htbp]
\caption{Evaluation results of scam classifiers.}
\centering
\begin{tabular}{l|cccc|cccc}
\hline
\multicolumn{1}{c|}{\multirow{2}{*}{Classifier}} & \multicolumn{4}{c|}{Cross-validation (balanced)}                                                     & \multicolumn{4}{c}{Test (imbalanced)}                                                              \\ \cline{2-9} 
\multicolumn{1}{c|}{}                            & \multicolumn{1}{c|}{accuracy} & \multicolumn{1}{c|}{precision} & \multicolumn{1}{c|}{recall} & F1    & \multicolumn{1}{c|}{accuracy} & \multicolumn{1}{c|}{precision} & \multicolumn{1}{c|}{recall} & F1   \\ \hline
List classifier                                   & \multicolumn{1}{c|}{1.0}    & \multicolumn{1}{c|}{1.0}       & \multicolumn{1}{c|}{1.0}  & 1.0     & \multicolumn{1}{c|}{0.99}    & \multicolumn{1}{c|}{1.0}     & \multicolumn{1}{c|}{0.99}  & 0.99    \\ \hline
\end{tabular}
\label{classifier_perf}
\end{table*}

\section{Background}
\label{sec:pre}
This section provides necessary background of the free giveaway scam disseminated through Twitter lists.
\subsection{Twitter List}
\label{sec:list}
Twitter List~\cite{me:twitterlist} is a function that allows users to create or follow an ad-hoc list of individual accounts they are interested in. With this function, people can create a list or follow an existing list created by others. After following a list, users can visit the list page and exclusively view tweets posted by the members in the list. The initial purpose of Twitter List is to facilitate advertisements of business and meet people' needs of following curated lists of influential people in their fields.

When creating a list, the creator can customize the list title and description and add other users to become a list member without requiring their explicit consent. Once users are added to the list, they will receive a notification prompting them to visit the list page. In this work, we observed that such a permission-less and pushing features of Twitter List had been abused by the free giveaway scammer to spread the scam to a large number of Twitter users.

\subsection{Free Giveaway Scams on Twitter Lists}
\label{sec:scam_giveaway}
The free giveaway is a fraudulent scheme that entices victims to transfer cryptocurrency to a designated account, promising to return double the amount or more. Scammers tend to compromise the social network account of celebrities~\cite{BrazenAttack} or create legitimate-looking websites to announce the giveaway scheme. Research have shown that several social network platforms such as YouTube, Telegram, and Twitter posts have been used to spread the URLs free giveaway websites~\cite{vakilinia2022cryptocurrency, xia20covidscams}. In this work, we identified a new distribution channel for spreading the free giveaway scam, which is the Twitter list. We show a sample Ethereum giveaway scam in Figure~\ref{fig:ether_scam} and describe the distribution process below.

\textbf{Step 1: Register a domain.} The scammer registers a legitimate-looking domain that will be used to publish the free giveaway scheme. To gain victims' trust, the scammer often injects popular names into the domain, including the name of blockchains, the name of blockchain companies or their founders. In the example, the \url{www.buterinofficial.com} is a domain injecting the name of the Ethereum founder, Vitalik Buterin.

\textbf{Step 2: Setup the free giveaway website.} The scammer then publish a website on the registered domain to announce the giveaway scheme. Figure~\ref{fig:ether_scam_site} shows the webpage hosted on the domain. The webpage claims to pay double the amount to participants who transfer more than 0.5 Ether to the address $0x6Da563…$. To further deceive victims, the webpage also periodically generates fake transactions indicating that people are actively participating, as shown in Figure~\ref{fig:ether_scam_txs}. Every 60 seconds a pair of fake transactions are generated, with a \textit{transfer-in} transaction displaying the participant (\textit{From}) and transferred amount (\textit{Quantity}), and a \textit{transfer-out} transaction displaying that the double amount is paid back to the participant.

\textbf{Step 3: Disseminate through Twitter lists.} Next, the scammer creates Twitter lists to announce the free giveaway in the list title and include the URL of the giveaway website in the description. Then the scammer adds Twitter users to the list and pushes a notification to them. Figure~\ref{fig:ether_scam_list} shows a sample list created by the scammer. The list title is "50000 ETH" and the description shows that the free giveaway is hosted on \textit{www.buterinofficial.com}. The scammer used a Twitter account impersonating the Ethereum founder and added $49$ members to the list. Finally, the scammer wait users to visit the list page. If users click the notification of being added to the list, they will be redirected to the list page.

\subsection{Existing work}
Below we introduce existing work on the free giveaway scam and defer the summary of other related work to Sec.~\ref{sec:related}.

Vakilinia et al.~\cite{vakilinia2022cryptocurrency}, Li et al.~\cite{xigao2023doublenothing}, and Xia et al.~\cite{xia20covidscams} have reported the free giveaway scam. Specifically, Vakilinia et al.~\cite{vakilinia2022cryptocurrency} manually collected 15 free giveaway scams disseminated in YouTube live streams. Xia et al.~\cite{xia20covidscams} developed a scam detection system to detect cryptocurrency scams by using "COVID-19" and "cryptocurrency" as keywords to collect data from Twitter posts, Telegram, etc. Their work reported 19 free giveaway scams. Li et al.~\cite{xigao2023doublenothing} also developed a scam detection system by using keywords to collect free giveaway domains from the Certificate Transparency Log (CT-Log)~\cite{me:ctlog} and reported 10K free giveaway scams. However, despite the contributions, these work have the following limitations. First, they all either focused on a few well-known distribution channels or utilized heuristic-based keywords to collect data to identify a scam, thus cannot uncover new scams disseminated in unknown distribution channels such as the Twitter List or scams that do not involve those heuristic-based keywords. Second, they all rely on manual effort to assess the collected data to detect a free giveaway scam, which is tedious and unscalable. 

%% file: text/system.tex
\section{New Scam Detection System}
\label{sec:system}
\begin{figure}[!ht]
  \centering
  \includegraphics[width=0.5\textwidth]{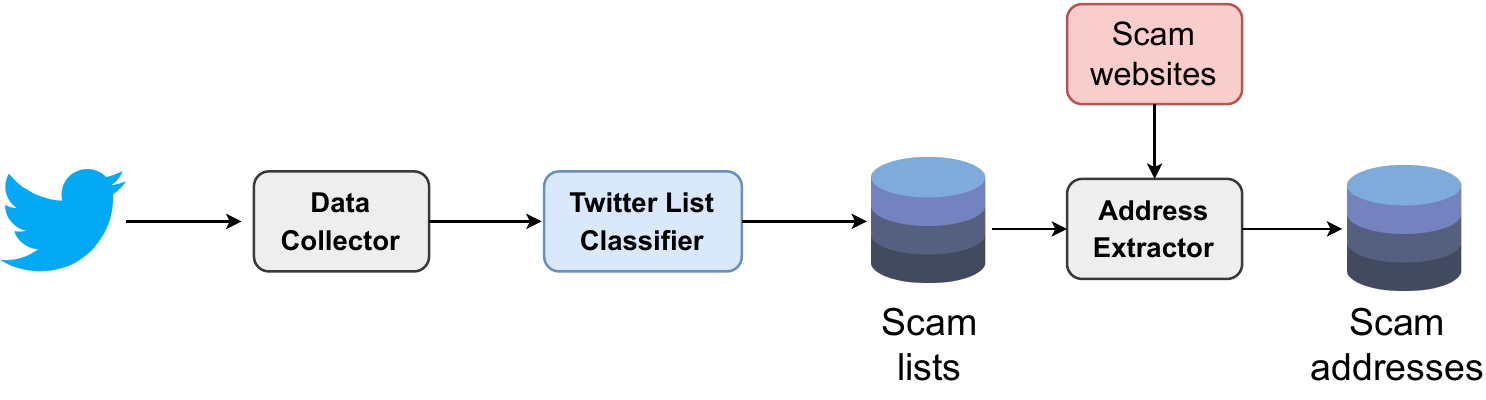}
  \caption{The detection pipeline of \textit{GiveawayScamHunter}.}%
  \label{fig:system}
\end{figure}

In this work, we aim to develop a fully automated scam detection to collect the free giveaway scam disseminated in the new distribution channel, Twitter List. Our detection system named \textit{GiveawayScamHunter} consists of three modules: a Data Collector, Scam Detector, and Address Extractor. The detection pipeline is depicted in Figure~\ref{fig:system}. Below we describe each module in detail.

\textbf{Data Collector:} To collect lists from Twitter, we leverage the Twitter Developer API~\cite{me:twitteracademicapi} to download lists from Twitter and retrieve relevant information of each list. Specifically, we utilize the List Lookup~\cite{me:listlookup} API to continuously search for new lists on Twitter. For each list, we record the list ID, title, description, creator, creation time, and list members.

\textbf{List Classifier:} Since the collected data may contain normal lists, it is necessary to classify them to detect free giveaway scam lists. Enlightened by the observation that a scam list typically announces the free giveaway in the list title and description, we propose to utilize Natural Language Processing (NLP) techniques to interpret the list title and description to detect the free giveaway scam. Specifically, we build a list classifier based on the text classification model in AllenNLP~\cite{AllenNLP}. Below we detail the training and evaluation process.

To train the list classifier, we first retrieve the title and description from the collected lists and preprocess them through the following steps: 1) we translate all text to English using Google Translation API~\cite{me:googletranslate}; 2) we remove duplicated records and filter out records that have less than 2 words. 3) we convert the emojis to their corresponding text representations~\cite{me:demoji} and replace the slangs with colloquial acronyms with their formal representations~\cite{me:slang}. After that we sample 2000 lists and manually label them with 3 annotators, each proficient in English and at least has an undergraduate college education. An external expert is involved to resolve inter-annotator disagreements. Through this annotation process, 27.9\% samples (or 532) were labeled as a scam. Given this distribution, we then randomly select 300 scam samples and 600 normal samples to construct the training data-set. We utilize oversampling to balance the training data-set by counting each scam sample twice. The remaining 1100 samples are used as the testing data-set. We perform 5-fold cross-validation to evaluate the training performance.

The cross-validation and evaluation results are presented in Table~\ref{classifier_perf}, including the metrics of accuracy, precision, recall, and F1-score. The results show that our list classifier can achieves a good performance of more than 99\% across all metrics in both the cross-validation and testing data-sets, demonstrating its generality on the whole data-set and the effectiveness in classifying and detecting free giveaway scams. 

\textbf{Address Extractor:} After detecting a scam list, the next step is to recognize and extract the cryptocurrency address used by the scammer for collecting funds. Our Address Extractor module relies on two steps to extract the scam address. First, we need to recognize the URL of free giveaway websites included in the list description and download the webpage hosted on the URL. To achieve this, we utilize URLExtract~\cite{me:urlextract} to recognize and extract URLs from the list description. After that we send a request to the URL to download the webpage. Second, we need to extract the cryptocurrency address embedded in the webpage. Based on the knowledge that each blockchain has its own unique address pattern, we utilize regular expressions to search cryptocurrency address from the downloaded webpage. For example, Bitcoin addresses are typically constructed by 26-35 alphanumeric characters that begin with a number 1 or 3 or a string '\textit{bc1}', and Ethereum addresses typically consist of 40 bytes hex numbers. We leverage this knowledge to compose regular expressions to extract the scam cryptocurrency address from the free giveaway scam webpage.

%% file: text/results_lists.tex
\section{Detection Results and Analysis}
\label{sec:list}
We implemented \textit{GiveawayScamHunter} in Python and deployed it over 1 year from June 2022 to June 2023. The overview of our collected data and detected scam lists are presented in Table~\ref{tab:overview}. In summary, we have collected \num{703576} Twitter lists, among which \num{95111} lists (or 13.5\%) were reported as a scam publishing the free giveaway scheme. Below we present a more in-depth analysis of the scam lists.

\begin{table}[!htbp]
\caption{Overview of collected lists and detected scam lists.}
\centering
\begin{tabular}{l|l|l|l}
\hline
Source         & Total  & Scam   & Ratio \\ \hline
Twitter lists  & \num{703576} & \num{95111} & 13.5\%     \\ \hline
\end{tabular}
\label{tab:overview}
\end{table}

\begin{table}[!htbp]
\caption{Distribution of creator accounts by the number of created scam lists.}
\centering
\begin{tabular}{lll}
\hline
\#Scam Lists & \#Accounts & Ratio \\ \hline
1                    & \num{77948}     & 92.1\%     \\
2                    & \num{6020}     & 7.1\%     \\
\textgreater{}2      & \num{686}      & 0.8\%      \\ \hline
\end{tabular}
\label{tab:list_creator_dist}
\end{table}

\begin{table*}[!htbp]
\caption{The profile summary of scam list creator accounts.}
\centering
\begin{tabular}{l|l|lll}
\hline
Account Status & \# (\%)     & \#Scam Lists (\%)  & \#Followers (\%)    & \#Tweets (\%)       \\ \hline
               &             & $1$ (90.6\%)      & 0  (29.5\%)    & 0  (33.0\%)    \\
Active         & \num{37182} (43.9\%) &           & $\leq5$  (63.6\%)  & $\leq5$  (50.4\%)  \\
               &             & $\geq2$ (9.4\%)   & $\leq10$  (72.3\%) & $\leq10$  (56.5\%) \\
               &             &                    & $>10$  (27.7\%)  & $>10$  (43.5\%)  \\ \hline
Suspended      & \num{47472} (56.1\%) & \multicolumn{3}{c}{N/A}                                          \\ \hline
\end{tabular}%
\label{tab:list_creator_profile}
\end{table*}

\subsection{Analysis of Free Giveaway Scam Lists} 
For the reported \num{95111} Twitter lists related to free giveaway scams, we aim to analyze the timeline trend of the scam on Twitter and understand the cryptocurrency categories targeted in the scam, as well as the profile of scam creator accounts and the scale of affected users.

\begin{figure}[!htbp]
\centering
 \subfloat[Lists created between 2019 and 2023.]{%
  \includegraphics[width=0.24\textwidth]{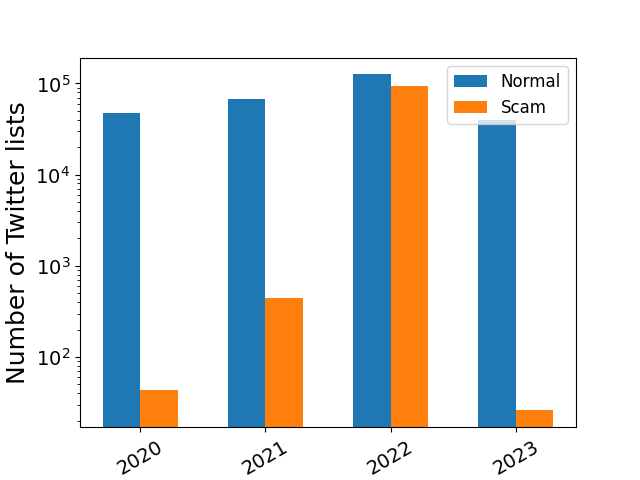}
  \label{fig:list_year}}
  \subfloat[Lists created between Jan. 2022 and Dec. 2022.]{%
  \includegraphics[width=0.24\textwidth]{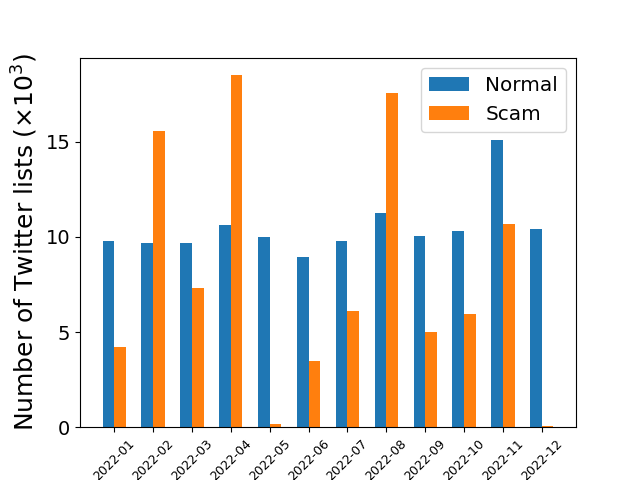}
   \label{fig:list_month}}
  \caption{Timeline trend of lists created on Twitter.}
\end{figure}
\textbf{Trend of scam lists:}
We retrieved the creation time from each list and grouped them into different time intervals (e.g., year and month). As the earliest scam list in our data-set dates back to 2020, we focused solely on lists created since then and show the distribution in Figure~\ref{fig:list_year}. First, it can be seen that since 2020, the number of normal lists created each year has exhibited linear growth, increasing from 47K in 2020 to 125K in 2022. In contrast, scam lists have experienced steep growth, rising from less than 500 in 2020 to 94K in 2022. Second, in 2023, the number of normal lists slightly decreased to 39K, while the number of scam lists significantly dropped below 30. Although this significant decrease in scam lists could be attributed to the shorter collection period in 2023, it still suggests that compared to 2022, free giveaway scammers are less active in publishing the scam on Twitter lists. Overall, our results indicate that most scam lists we collected were published in 2022, accounting for 99.5\% of all scam lists. 

Given that most scam lists were published in 2022, we then show the distribution of normal and scam lists published each month during that year in Figure~\ref{fig:list_month}. The result reveals that the scammers were highly active in February, April, and August, as evidenced by the creation of over 37K, 26K, and 18K scam lists, respectively. To gain further insights into this observed scam trend, we examined the cryptocurrency market trends~\cite{me:btcmonthlyprice, me:ethmonthlyprice} throughout 2022. Remarkably, we discovered a correlation between the scam and market trends. During the aforementioned months, cryptocurrency prices experienced surges, which seems to have stimulated scammers to publish scam lists on Twitter.
\begin{figure}[!htbp]
\centering
\includegraphics[width=0.43\textwidth]{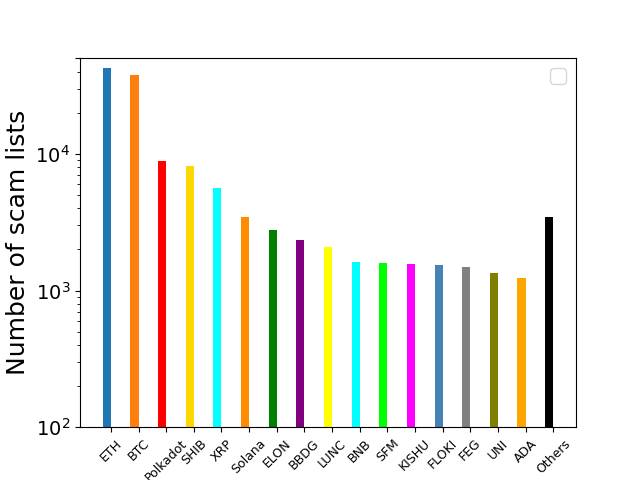}
\caption{Distribution of scam lists by cryptocurrency categories.}
\label{fig:list_crypto}
\end{figure}

\textbf{Cryptocurrency categories:} As described before, scam lists would typically announce which cryptocurrency will be given in the title. We thus extract the cryptocurrency name from the scam lists and present the distribution in Figure~\ref{fig:list_crypto}. As can be seen from the figure, we have identified more than \num{16} different cryptocurrency involved in the scam, including the well-known BTC (Bitcoin), ETH (Ethereum), XRP (Ripple), and BNB (Binance). Among them, being the most popular cryptocurrency in the market, ETH and BTC are targeted by most of the scam lists, as evidenced by over \num{43K} and \num{39K} lists. Following them are Polkadot, SHIB, and XRP, which were targeted by more than 5.6K lists.

\textbf{List creators:} Our next objective is to gain insights into scam creator accounts and analyze their Twitter profiles. To accomplish this, we first aggregate all scam lists based on the creator account and extract those unique accounts. In total, we identified \num{84654} distinct accounts. Table~\ref{tab:list_creator_dist} presents the distribution of scam accounts according to the number of lists they created. Notably, $92.1\%$ of accounts created only one scam list, and $7.1\%$ of accounts contributed to two scam lists. In contrast, only 0.8\% of accounts were associated with more than two scam lists. Subsequently, we analyze the profile of each scam creator account by retrieving the relevant public metrics such as the number of followers and tweets. We utilized the Twitter Users Lookup API~\cite{me:userlookup} and retrieved the public metrics in June 2023. The distribution result is presented in Table~\ref{tab:list_creator_profile}. Among the \num{84654} creator accounts, it was observed that $56.1\%$ of them had been suspended by Twitter, while the remaining $43.9\%$ remained active. Among the active accounts, we discovered that $29.5\%$ had no followers and $33\%$ had not posted any tweets. These observations strongly indicate that such accounts are likely spam accounts registered by the scammers. Moreover, it can be seen that $72.3\%$ of the active accounts have fewer than $10$ followers, and $56.5\%$ posted fewer than $10$ tweets.
\begin{table}[!htbp]
\caption{Profile of the top 10 accounts that created most of the scam lists. Usernames are anonymized for ethical considerations.}
\centering
\begin{tabularx}{0.43\textwidth}{lcccc}
\hline
 Username        & \#Scam Lists    & \#Followers       & \#Tweets  \\ \hline
 \textit{vm***}	 & \num{333}            & \num{856}	 & \num{815  }\\
 \textit{Aq***}  & \num{244}            & \num{40630}    & \num{3693 }\\
 \textit{ce***}  & \num{204}            & \num{63234}    & \num{755  }\\
 \textit{gu***}  & \num{107}            & \num{154721}   & \num{78669}\\
 \textit{ai***}  & \num{47 }            & \num{44}       & \num{218  }\\
 \textit{ce***}  & \num{47 }            & \num{96}       & \num{184  }\\
 \textit{m\_***}  & \num{45 }            & \num{80}       & \num{208  } \\
 \textit{m4***}  & \num{45 }            & \num{55}       & \num{4237 } \\
 \textit{om***}  & \num{43 }            & \num{664}	 & \num{505  } \\
 \textit{ts***}  & \num{43 }            & \num{1012}	 & \num{24808} \\\hline
\end{tabularx}
\label{tab:top_list_creator}
\end{table}

One intriguing observation in our analysis is that we identified ten accounts that were extremely active and had created more than \num{40} scam lists. Their detailed profiles are presented in Table~\ref{tab:top_list_creator}. We can see that account \textit{vm***} appeared as the most active one and published \num{333} scam lists. The account accumulated over 800 followers and tweets. Similarly, the accounts \textit{Aq***} and \textit{ce***} also demonstrated significant activity, with more than \num{200} scam lists created and over 40K followers accumulated. Among the ten accounts, \textit{gu***} stood out as the most popular one with more than \num{150}K followers. The account is a verified account and claims to be owned by a political leader. Upon analyzing the \num{107} scam lists created by this account, we found that all of them were publishing an "Ethereum and Bitcoin" giveaway hosted on \url{musk22-official.com}. Notably, all these lists were published between 00:43 and 00:48 am on February 27th, 2022. We suspect that the account was compromised during that time period, and the scammer exploited its established reputation and created many lists to lure victims.

\textbf{List members and followers:} We also conducted an analysis of the scale of members and followers in the scam lists. We retrieved the member and follower information from each scam list and present the distribution in Table~\ref{tab:list_member}. It can be seen that about 30\% of scam lists have less than 50 members, and 47\% of scam lists have 50 to 100 members. In contrast, only 23\% of them have more than 100 members. After being added to a list, a member can choose to become the follower by clicking the \textit{follow} button on the list page. In Table~\ref{tab:list_follower}, we show the distribution of scam lists by the number of followers. It can be seen that 56.8\% of scam lists have 0 followers, and 16.7\%, 6.3\%, and 3.6\% of them respectively have 1, 2, and 3 followers. The remaining 16.6\% of scam lists have more than 3 followers. Such a result indicates that most users were cautious and did not follow the list, however, there were a small fraction of users who showed interests and followed the scam lists. By retrieving the members included in all scam lists and counting the unique member accounts, we discovered that more than 3.7 million Twitter users have been affected by this type of scam.
\begin{table}[]
\caption{Distribution of scam lists by the number of members.}
\centering
\begin{tabular}{lll}
\hline
\#Members                          & \#Scam Lists                 & Ratio  \\ \hline
0 - 50                             & \num{29418} & 30.0\% \\
50 - 100                           & \num{46276} & 47.0\% \\
\textgreater{}100 & \num{22669} & 23.0\% \\ \hline
\end{tabular}
\label{tab:list_member}
\end{table}

\begin{table}[]
\caption{Distribution of scam lists by the number of followers.}
\centering
\begin{tabular}{lll}
\#Followers                        &                 &   \\ \hline
0                                  & \num{29418} & 30.0\% \\
1                                  & \num{46276} & 47.0\% \\
2                                  & \num{22669} & 23.0\% \\
3                                  & \num{46276} & 30.0\% \\
\textgreater{}3   & \num{22669} & 47.0\% \\ \hline
\end{tabular}
\label{tab:list_follower}
\end{table}

\textbf{Summary:} Overall, our analysis of scam lists has yielded several interesting observations. Firstly, we found a correlation between the surge in scam activity on Twitter and the fluctuations in the cryptocurrency market. Specifically, scammers tend to be more active during periods of cryptocurrency price increases. Secondly, the cryptocurrency targeted by these scam lists align with their popularity in the market, with the majority of lists focusing on the top two cryptocurrency. Furthermore, our investigation of the scam creator accounts reveals that most scammers avoid using the same account to publish multiple scam lists. Instead, they opt to register spam accounts specifically for this purpose, enabling them to disseminate scams more effectively. Lastly, when scammers gain access to a popular account, they create multiple scam lists to target a wider audience, leveraging the compromised account's reputation to instill trust and entice users to participate in the giveaway scam.

\subsection{Scam URLs}
Our next objective is to gain insights into the URLs of free giveaway websites included in the scam lists. We accomplish this by extracting URLs from the description of each scam list.
\begin{table}[!htbp]
\caption{Top 10 URLs appeared in the scam lists.}
\centering
\begin{tabularx}{.38\textwidth}{lcc}
\hline
URL            & \#Scam Lists          & Cryptocurrency  \\ \hline
\url{cryptoeventx2.com}   & \num{6708}            & BTC \& ETH      \\
\url{ethofficialx2.info}   & \num{5019}            & ETH             \\
\url{justevent.info}   & \num{4721}            & BTC \& ETH      \\
\url{czdrop.com}   & \num{4555}            & BTC             \\
\url{official-saylor.com}   & \num{4023}            & BTC \& ETH      \\
\url{coinx2.org}   & \num{2536}            & BTC \& ETH      \\
\url{xrpdouble.info}   & \num{2507}            & XRP             \\
\url{buterin.site}   & \num{2460}            & ETH             \\
\url{official-emusk.com}   & \num{2359}            & BTC \& ETH      \\
\url{babydoge.in}   & \num{2340}            & BABYDOGE        \\ \hline
\end{tabularx}
\label{tab:top_list_urls}
\end{table}

From the 95,111 scam lists, we extracted $327$ distinct URLs, indicating that most scam lists were publishing the same URL. We show in Table~\ref{tab:top_list_urls} the most frequent 10 URLs included in the scam lists and their targeted cryptocurrency. We can see that all of 10 URLs appeared in more than \num{2340} scam lists, with \url{www.cryptoeventx2.com} being the most frequent one included in 6.7K lists and targeting ETH and BTC. Among them, five URLs targeted both BTC and ETH, two targeted ETH only, and the remaining three targeted BTC, XRP, and BABYDOGE. Moreover, we also observed popular names injected in the domain name, including the targeted cryptocurrency name such as eth (ETH), xrp (XRP), and babydoge (BABYDOGE), and celebrities such as cz (CZ Zhao, CEO of Binance), buterin (Buterin Vitalik, Founder of Ethereum), and emusk (Elon Musk, CEO of Tesla). 

We also compared the free giveaway URLs collected in our work with that reported in existing studies~\cite{xia20covidscams, vakilinia2022cryptocurrency, xigao2023doublenothing}. Compared to \cite{xigao2023doublenothing}, only 17 (5.2\%) of our collected URLs were reported in their work and the other 310 (94.8\%) are uniquely discovered by our work. Compared to \cite{vakilinia2022cryptocurrency, xia20covidscams}, none of our discovered scam URLs were reported in their work.

%% file: text/results_profit.tex
\section{Scam Addresses and Transactions Analysis}
\label{sec:victim_profit}
So far, we have described our analysis of scam lists and the URLs of free giveaway. We then analyze the scam addresses used for collecting funds from victims and report the scale of victims and their financial loss.

\subsection{Scam Addresses}
As described in Sec.~\ref{sec:system}, our detection system consists of an \textit{Address Extractor} module utilizing the regular expressions to detect scam addresses from the webpage of free giveaway websites. By running \textit{Address Extractor} against the downloaded webpages, we have identified \num{121} unique scam addresses, as presented in Table~\ref{tab:profit}. It can be seen that the number of unique scam addresses is smaller than the number of distinct URLs, which can be explained as that multiple websites were actually advertising the same address, indicating that they are likely controlled by the same entity.

The table also shows our detected \num{121} scam addresses span over five different cryptocurrency, with 54 of them on Bitcoin and 48 on Ethereum that account for 92.6\% of all detected addresses. The remaining are 5 BNB addresses, 2 Cardano addresses, and 2 Ripple addresses.

\begin{table*}[!htbp]
\caption{Overview of free giveaway scam addresses on different blockchains and their collected profits}
\centering
\begin{tabular}{llcccccc}
\hline
Scam Type     & Cryptocurrency & \multicolumn{1}{c}{\begin{tabular}[c]{@{}c@{}}\#Scam \\  Address\end{tabular}} & \multicolumn{1}{c}{\begin{tabular}[c]{@{}c@{}}\#Txs btw. \\ Scammers\end{tabular}} & \multicolumn{1}{c}{\begin{tabular}[c]{@{}c@{}}\#Victim \\ Txs\end{tabular}} & \multicolumn{1}{c}{\begin{tabular}[c]{@{}c@{}}\#Victim \\  Address\end{tabular}} & \multicolumn{1}{c}{\begin{tabular}[c]{@{}c@{}}Total Victim\\ Cryptocurrency\\ Amount\end{tabular}} & \multicolumn{1}{c}{\begin{tabular}[c]{@{}c@{}}Total Victim \\  USD Value \\ (Min - Max)\end{tabular}} \\ \hline

              & Bitcoin (BTC)  & 64  &  0 &  72   & 54     & 5.39958            & 85.2K - 169.3K         \\
              & Ethereum (ETH) & 48  &  0 &  379  & 289       & 313.92          & 311.9K - 665.5K        \\
Free giveaway & Binance (BNB)  & 5   &  0 &  5    & 4    &  8.50714             & 1.7K - 3.0K            \\
              & Cardano (ADA)  & 2   &  0 &  10   & 11    & 42342               & 10.3K - 27.2K          \\
              & Ripple (XRP)   & 2   &  0 &  8   & 7    & 13030.9             & 4.0K - 7.0K            \\ \hline
\end{tabular}
\label{tab:profit}
\end{table*}
\subsection{Scam Victims and Financial Loss}
In most scam measurement studies, uncovering the scam victims and financial loss are challenging as the ground truth often relies on victim reports. However, thanks to the open nature of public blockchains allowing everyone to access the transactions recorded in the ledger, we are able to overcome this challenge by tracking transactions that involve the obtained scam address on each blockchain, which thus can assist us to identify the scam victims and estimate the financial loss. To obtain the transaction history of each scam address, we leverage the public API offered by blockchain explorers such as Etherscan~\cite{me:etherscan}, Bitcoin Tracker~\cite{me:btcscan}, BNB explorer~\cite{me:bnbexplorer}, BscScan~\cite{me:bscscan}, CardanoScan~\cite{me:adascan}, and XRPSCAN~\cite{me:xrpscan} to download transactions involving the scam addresses. Then we retrieve each transaction that transfers funds to the scam address and aggregate them together to estimate the scale of victims and financial loss. It is important to note that some transactions could be made by scammers themselves who intentionally transfer funds to themselves to create a deceptive appearance for potential victims. We thus conduct additional verification on the retrieved transactions by removing transactions between scam addresses.

\textbf{Scam transactions and victims:} We first discuss the analysis of the victim transactions. As can be seen from Table~\ref{tab:profit}, for the \num{121} free giveaway scam addresses, we found that there is no transaction between the extracted scam addresses and all \num{474} transactions from victim senders, with the majority of them (\num{379}) on the Ethereum blockchain. Then from the 474 victim transactions, we further identified 365 distinct victim addresses. With fewer victim addresses than the transactions, it implies that some victims were actually deceived more than once. Moreover, we can see that Ethereum and Bitcoin respectively have \num{289} and \num{54} victim addresses that contribute to 94\% of all victim addresses. Following them, Binance, Cardano, and Ripple blockchains have 4 to 11 victim addresses. 

\textbf{Financial loss:} Next, we discuss the analysis of financial loss we retrieved from the victim transactions. In Table~\ref{tab:profit}, we report both the total number of cryptocurrency transferred to scammers and their total USD value based on the time of the transfer. We can see that free giveaway scammers have made most of their profits from the Ethereum blockchain, with 313.92 Ethers collected and worth 311.9K to 665.5K USD. Following Ethereum, scammers stole more than 5.3 BTC and profited 85.2K to 169K USD on Bitcoin. From the reaming three blockchains, scammers have made a profit ranging from 1.7K to 27K USD. In total, free giveaway scammers discovered in our work have profited 413K to 872K USD. 

Overall, our analysis of scam transactions on the extracted scam addresses reveals several interesting facts. First, there is no transaction made between scam addresses, indicating that scammers seem not interested in playing additional tricks to deceive victims. Second, we found that some victims had been attacked more than once on the same blockchain and some of them were on two blockchains. Third, our analysis shows that the scam reported in our work have resulted in a financial loss of over 872K USD.

\subsection{Case Studies}
This section describes a case study on the most profitable free giveaway scam identified in our work.

\begin{figure}[!ht]
 \centering
 \includegraphics[width=0.48\textwidth]{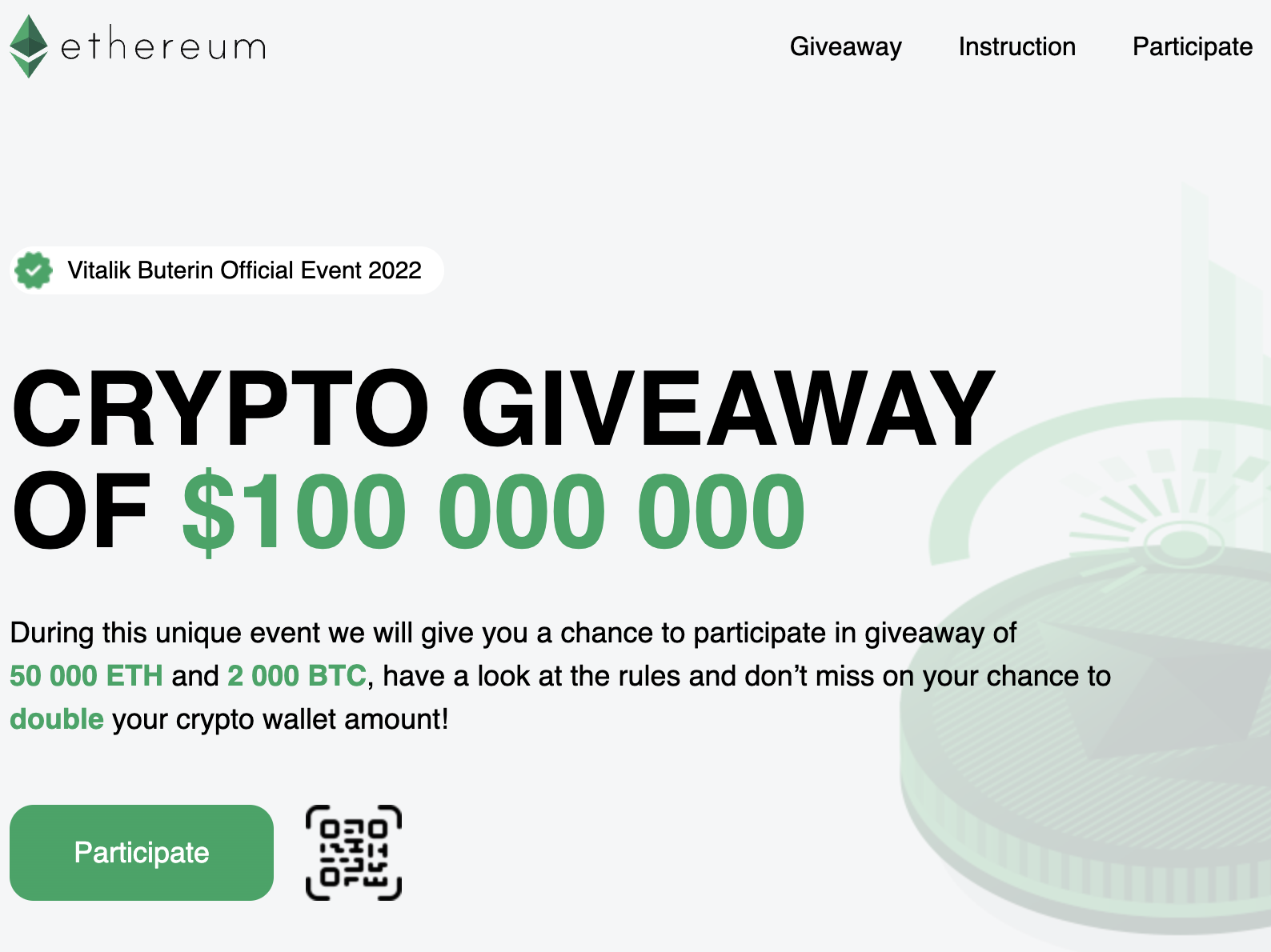}
 \caption{The webpage published in the most profitable free giveaway scam.}%
 \label{fig:case1}
\end{figure}
\textbf{Most profitable free giveaway scam:} Through the scam transaction analysis, we found that the most profitable free giveaway scam is an \textit{Ethereum and Bitcoin Giveaway} hosted on \url{https://merge-ethx2.info}. The scammer address is \hash{0xc4BE4a14d4E61b2DA5dd7eE64cbA8E85766dFD3d} on Ethereum, and \hash{13i79MkvCijL6MqjHL1ojyqmDREKHiYdbk} on Bitcoin. The scammer has collected more than $130.85$ Ethers from 4 victims that worth 130K to 277K USD. The biggest profit comes from transaction \hash{0x154a98e68828721ba1f0ca70ee2787d0ebb98ebf2926a30cb77cb25017ac8e80}. The transaction details show that the victim has transferred $130$ Ethers from a custodial Binance account to the scammer on Dec.24, 2022. In contrast, the scammer has not made any profits on the Bitcoin blockchain. Figure~\ref{fig:case1} shows the giveaway webpage created by the scammer. Compared to the sample giveaway website in Figure~\ref{fig:ether_scam}, there are several notable differences. First, this webpage is designed more professionally, showing well-formatted diagrams and text boxes. Second, on the webpage, there is a verified budget showing \textit{Vitalik Buterin Official Event 2022}. Third, the webpage also shows a picture of the Ethereum Founder \textit{Vitalik Buterin}. Lastly, at the bottom of the webpage, there are several badges showing the copyright and license information.

%% file: text/conclusion.tex
\section{Discussion}
\label{sec:discuss}
Below we discuss the disclosure of our collected scam and implications and limitations of our work.

\textbf{Responsible Disclosure:} We have reported the scam lists and associated spam accounts collected by our work to Twitter. In addition, we also disclosed the associated scam addresses to the blockchain community, including the blockchain explorer Etherscan~\cite{me:etherscan} and the blockchain security alerting platform HashDit~\cite{me:hashdit}.

\textbf{Implications:} The free giveaway scam is an investment scam which exploits victims' desire of making profits quickly. Due to the anonymity and in-reversibility of blockchain, the victims' financial loss is difficult to recover. It is thus important to raise people's awareness of such a scam. In addition, it is necessary for social network platforms to timely detect cryptocurrency scams and prevent their dissemination's. Our detection system can serve as the first step in solving this problem. Besides, our detection system can also benefit other services such as blockchain explorers, digital wallets, and exchange services, allowing them to flag the detected scam address and terminate the withdrawals to protect victims.

\textbf{Limitations:} Our work entails the following limitations. First, our data collection was started in the middle of 2022, so the reported trend before that time could be biased as some scam lists and spam accounts may have already been suspended on Twitter. Second, the scale of scam victims and profits were estimated based on the transactions involving the detected scam address, though transactions between the detected scam addresses have been removed, the transactions could still be made by other unknown scam addresses. Hence our result should be interpreted as the upper-bound of the actual financial loss of victims.

\input{text/related_work.tex}
\section{Conclusion}
\label{sec:conclusion}
This paper presents a large scale analysis of the free giveaway scams disseminated on Twitter lists. By running \textit{GiveawayScamHunter} in 1 year, we collected tens of thousands of free giveaway scam lists and discovered hundreds of giveaway URLs and scam addresses. Our work sheds light on the dissemination strategies and behaviors of the scammers on Twitter, as well as the scale of scam victims and financial loss, emphasizing the urgent need to prevent the dissemination of scams on social networks and protect users from falling into such a fraudulent activity.

%% file: text/related_work.tex
\section{Related Work}
\label{sec:related}
In the existing literature, there are various work studying cryptocurrency scams, including Ponzi Schemes~\cite{kell2021forsage, bian2021image, xia20covidscams, bartoletti2020dissecting, bartoletti2018data, chen2018detecting}, phishing scams~\cite{chen2020phishing, badawi2020automatic}, scam tokens~\cite{gao2020tracking, xia21scams}, fraudulent Initial Coin Offering~\cite{phua2022don, chiu2022using, liebau2019crypto, zetzsche2017ico}, fake exchange scams~\cite{xia2020characterizing}, and giveaway scams~\cite{xia20covidscams, vakilinia2022cryptocurrency, xigao2023doublenothing}. Specifically, ~\cite{xia21scams} proposed a machine-learning-based approach to detect scam tokens listed on UniSwap and discovered more than 10K scam tokens. ~\cite{xia2020characterizing} implemented an automated scam detection framework on exchange domains and discovered 1595 scam domains and 300 fake exchange apps. ~\cite{wu2020phishers} proposed an approach to detect phishing scams on Ethereum through a novel embedding algorithm to extract the features of cryptocurrency addresses, including transaction amount and timestamp, which can identify phishing addresses.

Vakilinia et al.~\cite{vakilinia2022cryptocurrency}, Li et al.~\cite{xigao2023doublenothing}, Xia et al.~\cite{xia20covidscams} have reported the free giveaway scam and some of them also developed a scam detection system. Compared to them, our work has several notable differences. First, our system \textit{GiveawayScamHunter} does not use heuristic keywords to find the free giveaway URLs. Second, our system \textit{GiveawayScamHunter} is fully automated and does not rely on manual effort to assess the collected data to determine the free giveaway scam. Third, our work presents a unique analysis of the scammers' behaviors on Twitter and has revealed their dissemination strategies.

%% file: paper.bbl
\begin{thebibliography}{10}
\providecommand{\url}[1]{#1}
\csname url@samestyle\endcsname
\providecommand{\newblock}{\relax}
\providecommand{\bibinfo}[2]{#2}
\providecommand{\BIBentrySTDinterwordspacing}{\spaceskip=0pt\relax}
\providecommand{\BIBentryALTinterwordstretchfactor}{4}
\providecommand{\BIBentryALTinterwordspacing}{\spaceskip=\fontdimen2\font plus
\BIBentryALTinterwordstretchfactor\fontdimen3\font minus
  \fontdimen4\font\relax}
\providecommand{\BIBforeignlanguage}[2]{{%
\expandafter\ifx\csname l@#1\endcsname\relax
\typeout{** WARNING: IEEEtran.bst: No hyphenation pattern has been}%
\typeout{** loaded for the language `#1'. Using the pattern for}%
\typeout{** the default language instead.}%
\else
\language=\csname l@#1\endcsname
\fi
#2}}
\providecommand{\BIBdecl}{\relax}
\BIBdecl

\bibitem{kell2021forsage}
T.~Kell, H.~Yousaf, S.~Allen, S.~Meiklejohn, and A.~Juels, ``Forsage: Anatomy
  of a smart-contract pyramid scheme,'' \emph{arXiv preprint arXiv:2105.04380},
  2021.

\bibitem{bian2021image}
L.~Bian, L.~Zhang, K.~Zhao, H.~Wang, and S.~Gong, ``Image-based scam detection
  method using an attention capsule network,'' \emph{IEEE Access}, vol.~9, pp.
  33\,654--33\,665, 2021.

\bibitem{xia20covidscams}
P.~Xia, H.~Wang, X.~Luo, L.~Wu, Y.~Zhou, G.~Bai, G.~Xu, G.~Huang, and X.~Liu,
  ``Don’t fish in troubled waters! characterizing coronavirus-themed
  cryptocurrency scams,'' in \emph{2020 APWG Symposium on Electronic Crime
  Research (eCrime)}, 2020, pp. 1--14.

\bibitem{bartoletti2020dissecting}
M.~Bartoletti, S.~Carta, T.~Cimoli, and R.~Saia, ``Dissecting ponzi schemes on
  ethereum: identification, analysis, and impact,'' \emph{Future Generation
  Computer Systems}, vol. 102, pp. 259--277, 2020.

\bibitem{bartoletti2018data}
M.~Bartoletti, B.~Pes, and S.~Serusi, ``Data mining for detecting bitcoin ponzi
  schemes,'' in \emph{2018 Crypto Valley Conference on Blockchain Technology
  (CVCBT)}.\hskip 1em plus 0.5em minus 0.4em\relax IEEE, 2018, pp. 75--84.

\bibitem{chen2018detecting}
W.~Chen, Z.~Zheng, J.~Cui, E.~Ngai, P.~Zheng, and Y.~Zhou, ``Detecting ponzi
  schemes on ethereum: Towards healthier blockchain technology,'' in
  \emph{Proceedings of the 2018 world wide web conference}, 2018, pp.
  1409--1418.

\bibitem{chen2020phishing}
W.~Chen, X.~Guo, Z.~Chen, Z.~Zheng, and Y.~Lu, ``Phishing scam detection on
  ethereum: Towards financial security for blockchain ecosystem.'' in
  \emph{IJCAI}, vol.~7, 2020, pp. 4456--4462.

\bibitem{badawi2020automatic}
E.~Badawi, G.-V. Jourdan, G.~Bochmann, and I.-V. Onut, ``An automatic detection
  and analysis of the bitcoin generator scam,'' in \emph{2020 IEEE European
  Symposium on Security and Privacy Workshops (EuroS\&PW)}.\hskip 1em plus
  0.5em minus 0.4em\relax IEEE, 2020, pp. 407--416.

\bibitem{gao2020tracking}
B.~Gao, H.~Wang, P.~Xia, S.~Wu, Y.~Zhou, X.~Luo, and G.~Tyson, ``Tracking
  counterfeit cryptocurrency end-to-end,'' \emph{Proceedings of the ACM on
  Measurement and Analysis of Computing Systems}, vol.~4, no.~3, pp. 1--28,
  2020.

\bibitem{xia21scams}
\BIBentryALTinterwordspacing
P.~Xia, H.~Wang, B.~Gao, W.~Su, Z.~Yu, X.~Luo, C.~Zhang, X.~Xiao, and G.~Xu,
  ``Trade or trick? detecting and characterizing scam tokens on uniswap
  decentralized exchange,'' \emph{Proc. ACM Meas. Anal. Comput. Syst.}, vol.~5,
  no.~3, dec 2021. [Online]. Available: \url{https://doi.org/10.1145/3491051}
\BIBentrySTDinterwordspacing

\bibitem{xia2020characterizing}
P.~Xia, H.~Wang, B.~Zhang, R.~Ji, B.~Gao, L.~Wu, X.~Luo, and G.~Xu,
  ``Characterizing cryptocurrency exchange scams,'' \emph{Computers \&
  Security}, vol.~98, p. 101993, 2020.

\bibitem{vakilinia2022cryptocurrency}
I.~Vakilinia, ``Cryptocurrency giveaway scam with youtube live stream,'' in
  \emph{2022 IEEE 13th Annual Ubiquitous Computing, Electronics \& Mobile
  Communication Conference (UEMCON)}.\hskip 1em plus 0.5em minus 0.4em\relax
  IEEE, 2022, pp. 0195--0200.

\bibitem{xigao2023doublenothing}
X.~Li, A.~Yepuri, and N.~Nikiforakis, ``Double and nothing: Understanding and
  detecting cryptocurrency giveaway scams,'' 2023.

\bibitem{me:twitterlist}
``About twitter lists,''
  \url{https://help.twitter.com/en/using-twitter/twitter-lists}.

\bibitem{BrazenAttack}
\BIBentryALTinterwordspacing
K.~C. S.~Frenkel, N.~Popper and D.~E. Sanger, ``A brazen online attack targets
  v.i.p. twitter users in a bitcoin scam,'' 2020. [Online]. Available:
  \url{https://www.nytimes.com/2020/07/15/technology/twitter-hack-billgates-elon-musk.html,
  2020.}
\BIBentrySTDinterwordspacing

\bibitem{me:ctlog}
``Crt.sh,'' \url{https://crt.sh/}, Retrieved Feb, 5, 2023.

\bibitem{me:twitteracademicapi}
``Twitter api for academic research | products,''
  \url{https://developer.twitter.com/en/products/twitter-api/academic-research}.

\bibitem{me:listlookup}
``List lookup,''
  \url{https://developer.twitter.com/en/docs/twitter-api/lists/list-lookup/api-reference/get-users-id-owned_lists}.

\bibitem{AllenNLP}
\BIBentryALTinterwordspacing
M.~Gardner, J.~Grus, M.~Neumann, O.~Tafjord, P.~Dasigi, N.~Liu, M.~Peters,
  M.~Schmitz, and L.~Zettlemoyer, ``Allennlp: A deep semantic natural language
  processing platform,'' 2018. [Online]. Available:
  \url{https://arxiv.org/abs/1803.07640}
\BIBentrySTDinterwordspacing

\bibitem{me:googletranslate}
``Cloud translation api,''
  \url{https://cloud.google.com/translate/docs/reference/rest}.

\bibitem{me:demoji}
``demoji,'' \url{https://pypi.org/project/demoji}.

\bibitem{me:slang}
``Internet \& text slang dictionary,''
  \url{https://www.noslang.com/dictionary}.

\bibitem{me:urlextract}
``urlextract - pypi,'' \url{https://pypi.org/project/urlextract}.

\bibitem{me:btcmonthlyprice}
``Bitcoin monthly prices,''
  \url{https://finance.yahoo.com/quote/BTC-USD/history?period1=1638835200&period2=1675728000&interval=1mo&filter=history&frequency=1mo&includeAdjustedClose=true}.

\bibitem{me:ethmonthlyprice}
``Ethereum monthly prices,''
  \url{https://finance.yahoo.com/quote/ETH-USD/history?period1=1638835200&period2=1675728000&interval=1mo&filter=history&frequency=1mo&includeAdjustedClose=true}.

\bibitem{me:userlookup}
``Users lookup,''
  \url{https://developer.twitter.com/en/docs/twitter-api/users/lookup/api-reference/get-users}.

\bibitem{me:etherscan}
``Etherscan: Ethereum (eth) blockchain explorer,'' \url{https://etherscan.io/
  }.

\bibitem{me:btcscan}
``Blockchain explorer - bitcoin tracker,''
  \url{https://www.blockchain.com/explorer}.

\bibitem{me:bnbexplorer}
``Bnb chain explorer,'' \url{https://explorer.bnbchain.org }.

\bibitem{me:bscscan}
``Bscscan: Bnb smart chain explorer,'' \url{https://bscscan.com/ }.

\bibitem{me:adascan}
``Cardanoscan - cardano explorer,'' \url{https://cardanoscan.io/}.

\bibitem{me:xrpscan}
``Xrpscan: Xrp ledger explorer,'' \url{https://xrpscan.com/}.

\bibitem{me:hashdit}
``Hashdit - securing bnb chain,'' \url{https://www.hashdit.io/en }.

\bibitem{phua2022don}
K.~Phua, B.~Sang, C.~Wei, and G.~Y. Yu, ``Don't trust, verify: The economics of
  scams in initial coin offerings,'' \emph{Available at SSRN 4064453}, 2022.

\bibitem{chiu2022using}
T.~Chiu, V.~Chiu, T.~Wang, and Y.~Wang, ``Using textual analysis to detect
  initial coin offering frauds,'' \emph{Journal of Forensic Accounting
  Research}, vol.~7, no.~1, pp. 165--183, 2022.

\bibitem{liebau2019crypto}
D.~Liebau and P.~Schueffel, ``Crypto-currencies and icos: Are they scams? an
  empirical study,'' \emph{An Empirical Study (January 23, 2019)}, 2019.

\bibitem{zetzsche2017ico}
D.~A. Zetzsche, R.~P. Buckley, D.~W. Arner, and L.~F{\"o}hr, ``The ico gold
  rush: It's a scam, it's a bubble, it's a super challenge for regulators,''
  \emph{University of Luxembourg Law Working Paper}, no.~11, pp. 17--83, 2017.

\bibitem{wu2020phishers}
J.~Wu, Q.~Yuan, D.~Lin, W.~You, W.~Chen, C.~Chen, and Z.~Zheng, ``Who are the
  phishers? phishing scam detection on ethereum via network embedding,''
  \emph{IEEE Transactions on Systems, Man, and Cybernetics: Systems}, vol.~52,
  no.~2, pp. 1156--1166, 2022.

\end{thebibliography}
